\documentclass[a4paper,superscriptaddress,prl,showkeys,twocolumn]{revtex4}

\usepackage[english]{babel}
\usepackage{graphicx}
\usepackage{amsmath}
\usepackage{epstopdf}

\begin{document}

\title{Inhomogeneous and anisotropic Universe and apparent acceleration}

\author{G. Fanizza}
\email{giuseppe.fanizza@ba.infn.it}
\affiliation{Dipartimento di Fisica, Universit\`a di Bari, Via G. Amendola 173, 70126, Bari, Italy}
\affiliation{Istituto Nazionale di Fisica Nucleare, Sezione di Bari, Bari, Italy}

\author{L. Tedesco}
\email{luigi.tedesco@ba.infn.it}
\affiliation{Dipartimento di Fisica, Universit\`a di Bari, Via G. Amendola 173, 70126, Bari, Italy}
\affiliation{Istituto Nazionale di Fisica Nucleare, Sezione di Bari, Bari, Italy}

\date{\today}

\begin{abstract}
In this paper we introduce a LTB-Bianchi I (plane symmetric) model of Universe. We study and solve Einstein field equations. We investigate the effects of such model of Universe in particular these results are important in understanding the effect of the combined presence of an inhomogeneous and anisotropic Universe. The observational magnitude-redshift data deviated from UNION 2 catalog has been analyzed in the framework of this LTB-anisotropic Universe and the fit has been achieved without the inclusion of any dark energy.
\end{abstract}
\maketitle
\section{I. INTRODUCTION}
A very important assumption of the standard model of cosmology ($\Lambda \text{CDM}$) is based on the homogeneous and isotropic Friedmann-Lemaitre-Robertson-Walker solutions of Einstein's equations. The homogeneity and the isotropy are considered on large scale in the Universe.
The Universe is not isotropic or spatially homogeneus on local scales.

The question of whether the Universe is homogeneous and isotropic is of fundamental importance to cosmology, but we have not decisive answers. On the other hand neither observations of luminosity distance combinated with galaxy number counts nor isotropic Cosmic Microwave Background Radiation  are able to say if the Universe is spatially homogeneous and isotropic. 

The fundamental question consists in a simple observation: this geometry is the only that is able to explain and to be compatible with experimental data?  Are we sure that the assumption of homogeneity and isotropy is a logical and  comforting  way of thinking, or better is it an {\it a-priori} assumption?
\\
This is a pertinent question because we need that more $96\%$ of the content of our Universe must be dark (energy and matter) in order to have a compatible model with observations. The solution of the dark energy puzzle is the keystone of the modern cosmology.

Are there observables that can prove the Universe is homogeneous and isotropic on large scales? Very interesting studies has been done in this direction \cite{{Ehlers:1966ad},{Clarkson:1999yj},{Clarkson:2010uz}}.

The $\Lambda$CDM model of the Universe is remarkably successfull, but we have important tensions between the model and the experimental data \cite{{percival},{devega}}. On the other hand dark energy is the biggest puzzle in cosmology. 
There are many papers with more detailed discussions about dark energy, that are outside the scope of this paper, see for example \cite{blanchard,clifton} and references therein.  There are many reasons that consider the $\Lambda$CDM model full of theoretical problems \cite{copeland}, one is that $\Lambda$ has a value absurdly small in quantum physics. Moreover we cannot expect that dark energy will have in future locally observable effects.
\\
\\
The Cosmic Microwave Background (CMB) has high isotropy and this is considered as a strong evidence of the homogeneity and isotropy of the Universe, that is to say the Universe is well described by means of a Friedmann-Lemaitre-Robertson-Walker (FLRW) model. The main indication for this model is due to the Ehlers, Geren and Sachs theorem (EGS) \cite{Ehlers:1966ad} in 1968. This theorem is due to an earlier paper of Tauber and Weinberg \cite{Tauber:1961lbq} in 1961. In EGS theorem we consider the observers in an expanding Universe, dust Universe measures isotropic CMB and this implies that FLRW metric is valid and the cosmological principle is also valid. This theorem is important because it permits to have the homogeneity and isotropy not from experimental measurements of the isotropy of the Universe but from  the CMB. But as we will discuss later, CMB radiation have small anisotropies with $10^{-5}$ of amplitude!
\\
\\
As regards the homogeneity of the Universe 
it is important to note that the mass density of the Universe is not inhomogeneous on scales much smaller than the Hubble radius, in other terms  the homogeneity is not true at all orders but we can assume to be valid on distance greater that 100 Mpc. Many papers indicate this feature,  see for example \cite{sylos1} (and references therein), where the author indicates evidences that galaxy distribution is spatially inhomogenous for $r < 100 \, \text{Mpc/h}$.
\\
The strong interest in inhomogenous cosmological models, in particular the so called Lemaitre-Tolman-Bondi (LTB) model \cite{lemaitre,tolman,bondi} (for more details see \cite{krasinsky} and references therein), that represents a spherically symmetric exact solution to the Einstein's equations with pressureless  ideal fluid, is due to the its simplicity and it is very useful. In fact it allows for studies of inhomogeneities that cannot be analyzed as perturbative deviations from FLRW and it permits to evaluate the effect of inhomogeneities In particular it has been studied that LTB models without dark energy can fit observed data. 
\\
The high precision cosmology is able to understand by more details our  study about Universe. When we consider the isotropy of CMB we must not forget that this is not sufficient to say that our region of space is isotropic \cite{maartens2011}.

We have two very important observational evidences showing that we don't have exact isotropy \cite{russel}. Both evidences may be caused by an anisotropic phase during the evolution of our Universe in other terms the existence of anomalies in CMB suggests the presence of an anomalous plane-mirroring symmetry on large scales \cite{{gurzadyan1},{gurzadyan2}}. The same anomalous features in seven-years WMAp data and Planck data seems to suggest that our Universe could be non-isotropic .

The first is the presence of small anisotropy deviations as regards the isotropy of the CMB. In fact we have small anisotropies with $10^{-5}$ amplitude.

The second is connected with the presence of large angle anomalies \cite{bennet}. These anomalies can be considered in 4 families: 

1) the alignment of quadrupole and octupole moments \cite{{land},{Ralston:2003pf},{de OliveiraCosta:2003pu},{Copi:2003kt}}; 

2) the large scale asymmetry \cite {{Eriksen:2003db},{Hansen:2004vq}};
 
3) the very strange cold spot \cite{Vielva:2003et}; 
 
4) the low quadrupole moment of the CMB, that is very important because it may indicate an ellipsoidal - Bianchi type I anisotropic evolution of the Universe \cite{{barrow},{campanelli1},{campanelli2},{berera}}. This is due to the fact that the low quadrupole moment is suppressed at large scale and this suppression cannot be explained by the common cosmological model. 
\\

Some years ago it has been shown  \cite {martinez} that if we start with a FLRW Universe, it is possible to have small deviations from homogeneity and isotropy taking into account small deviations in the CMB. In particular if we consider an homogeneous and anisotropic Universe, the small quadrupole anisotropy in CMB implies a very small anisotropy in the Universe. Next, general results have been established \cite{maartellisstoe1,maartellisstoe2}, in which the authors does not assume a priori homogeneity and they found that small anisotropies in CMB imply that the Cosmo is not exactly FLRW but it is almost FLRW. Limits on anisotropy and inhomogeneity can be found starting from CMB. 
\\
The cosmological model that takes into account all these and stimulates many interest is the "anisotropic Bianchi type I model" that can be an intriguing alternative to the standard model FLRW, in which small deviations from the isotropy is able to explain the anisotropies and the anomalies in the CMB.
\\
The anisotropy considered in this work might be interpreted as an imprinting, a primordial relic of an early anisotropy that appears in the context of a multi- dimensional cosmological model of unified string theories.

In this paper our goals are to study an anisotropic and inhomogeneous model of the Universe.  In particular we introduce a new approach to a Universe in which inhomogeneities and anisotropies coexist, therefore we study  in order to obtain the relative Einstein's equations. These models of Universe inhomogeneous and anisotropic has been studied in different physical situations, as the role of the diffusion forces in governing the large-scale dynamics of inhomogeneous and anisotropic Universe \cite{shogin}.

The Supernovae observations are good tests about the structure of the space-time on different scales.  This is a very important point, in fact some years ago Zel'dovich \cite{zeld} studied the importance of the effects of the inhomogeneities on light propagation and also in the next years \cite{bertotti, dash,gunn,kant,dyer,weinberg,ellis1}. 
To check this model we calculate the luminosity distance in order to compare theoretical approach with experimental data. We explain the acceleration of the Universe without invoking the presence of a cosmological constant  or dark energy.
\\
\\
The structure of this paper is the following. In the next Section we calculate the metric for this LTB-Bianchi I model of the Universe. In Section III after providing the calculation of  various symbols we write the Einstein's equations taking into account this geometry. Section IV is dedicated to calculate the luminosity distance and in Section V we compare the theoretical data with experimental data.  Finally the discussion and conclusion are summarized in Section VI.
\\
\\
{\section {II. LTB-Bianchi 1 metric}}   
In order to find the anisotropic-LTB metric, let us start with a Bianchi type I space-time metric, spatially homogeneous, descripted by the metric 
\begin{equation}
\label{bianchi1}
ds^2= dt^2 -a^2(t) \, (dx^2 +dy^2) - b^2(t) \, dz^2
\end{equation}
with two expansion parameters $a$ and $b$ that are the scale factors normalized in order that $a(t_0)=b(t_0)=1$ and $t_0$ present cosmic time. The metric (\ref{bianchi1}) considers the xy-plane as a symmetry plane. 
 To our aim we write the Bianchi type I metric in polar coordinate $(x=r \, \text{sin} \theta \, \text{cos} \phi, y=r \,\text{sin} \theta \, \text{sin} \phi, z=r\, \text{cos} \theta)$:
\begin{eqnarray}
\label{bianchi1polare}
ds^2 & = & dt^2 - [a^2(t)  {\text{sin}}^2 \theta + b^2(t)  {\text{cos}}^2 \theta ] \, dr^2 -r^2 \, [a^2(t)  {\text{cos}}^2 \theta  \nonumber 
\\
& &
+ b^2(t) \, {\text{sin}}^2 \theta] \, d \theta^2 -2r \, [a^2(t) - b^2(t)] \text{sin} \theta \, \text{cos} \theta \, dr \, d \theta  \nonumber 
\\
& & 
- r^2 \, a^2(t) \, {\text{sin}}^2 \theta \, d \phi^2.
\end{eqnarray} 
In order to have a LTB-Bianchi I metric, we make the following substitutions 
\begin{equation}
\label{subst1}
r \, a(t) \rightarrow A_\rVert(r,t)\equiv A_\rVert \;  
\end{equation}
\begin{equation}
\label{subst2}
r \, b(t) \rightarrow A_\perp(r,t) \equiv A_\perp.
\end{equation}
In this way it is possible to obtain the general LTB-Bianchi I metric in polar coordinate, observing that $a(t)=A'$ and $2\,r '\, a^2(t) =(A^2_{\rVert})'$ where $' \equiv \partial/ \partial r$, we have:
\begin{eqnarray}
\label{LTB-bianchi1}
ds^2 &=&  dt^2 - ({A'^2_\rVert} \, {\text{sin}}^2 \theta + {A'^2_\perp} \, {\text{cos}}^2 \theta ) dr^2 - (A^2_\rVert \, {\text{cos}}^2 \theta + \nonumber
\\
& & 
+ A^2_\perp \, {\text{sin}}^2) \, d \theta^2 - ({A^2_\rVert}' - {A^2_\perp}') \, \text{sin} \theta \, \text{cos} \theta \,dr \, d \theta + \nonumber\\
& & 
- A^2_\rVert \, {\text{sin}}^2 \theta \, d \phi^2 .
\end{eqnarray} %
It is important to observe that the eq.~(\ref{LTB-bianchi1}) brings back to known cases: 
\begin{equation}
\begin{cases}
\label{eq:Bianchicase}
A_\rVert(r,t)=r\,a(t)\,\,\,\text {and}\,\,\,A_\perp(r,t)=r\,b(t), &\quad\text{Bianchi I}\\
A_\rVert(r,t)=A_\perp(r,t), &\quad\text{LTB}\\
A_\rVert(r,t)=A_\perp(r,t)=r\,a(t) &\quad\text{FRW}.
\end{cases}
\end{equation}%
Therefore the metric (\ref{LTB-bianchi1}) is a non-homogeneous metric with axial symmetry, that is simple referable to pure homogeneous or pure isotropic case.
\\
Let us define the following quantity 
\begin{equation}
\label{epsilon}
\epsilon(r,t)=A_\perp - A_\rVert
\end{equation}
that represents the degree of anisotropy of the Universe. 
From the definition of $\epsilon$ we obtain:
\begin{subequations}
\allowdisplaybreaks
\begin{align}
\label{eq:aniDer}
A'_\perp=&\,A'_\rVert+\epsilon'\\
\label{eq:aniDerSquared}
{A'_\perp}^2=&\,{A'_\rVert}^2+\epsilon'^2+2\,A'_\rVert\,\epsilon'\\
\label{eq:aniSquared}
A_\perp^2=&\,A_\rVert^2+\epsilon^2+2\,A_\rVert\,\epsilon\\
\label{eq:aniSquaredDer}
\left(A_\perp^2\right)'=&\left(A_\rVert^2\right)'+\left(\epsilon^2\right)'+2\,A'_\rVert\,\epsilon+2\,A_\rVert\,\epsilon'.
\end{align}
\end{subequations}
Let us introduce these relations in the metric (\ref{LTB-bianchi1}) in order to show it as a function of $\epsilon$ and $A_{\rVert}$ (or $\epsilon$ and $A_{\perp}$). Putting all togheter we have
\begin{eqnarray}
\label{eq:LTBplusAnisotropic2}
ds^2&=&dt^2-\left[ {A'_\rVert}^2+(\epsilon'^2+2\,A'_\rVert\,\epsilon') \cos^2 \theta \right]dr^2+\nonumber\\&-&\left[ {A_\rVert}^2+(\epsilon^2+2\,A_\rVert\,\epsilon) \sin^2 \theta \right] d\theta^2+\nonumber\\&+&\left[ \left(\epsilon^2\right)'+2\,A'_\rVert\,\epsilon+2\,A_\rVert\,\epsilon' \right]\sin \theta \cos \theta\,dr d\theta +\nonumber \\&-&{A_\rVert}^2 \sin^2 \theta d\phi^2\equiv\nonumber\\&\equiv& \left(g_{\rVert\mu \nu}^\text{(LTB)}+\Delta g_{\rVert\mu \nu}^{(AN)}\right)dx^\mu dx^\nu
\end{eqnarray}
with our metric given by 
\begin{equation}
\label{metric}
g_{\mu \nu}\equiv g_{\rVert \mu \nu}^\text{(LTB)}+\Delta g_{\rVert \mu \nu}^{(AN)}
\end{equation}
where:
\begin{subequations}
\allowdisplaybreaks
\begin{align}
\label{eq:pert10}
g_{\rVert \mu \nu}^\text{(LTB)}=&\left(
\begin{matrix}
1&0&0&0\\
0&-{A'_\rVert}^2&0&0\\
0&0&-A_\rVert^2&0\\
0&0&0&-A_\rVert^2 \sin^2 \theta
\end{matrix}
\right)\\
\label{eq:pert111}
\Delta g_{\rVert11}^{(AN)}=&-\cos^2 \theta\,\left(\epsilon'^2+2\,A'_\rVert\,\epsilon'\right)\\
\label{eq:pert112}
\Delta g_{\rVert12}^{(AN)}=&\frac{\sin{2 \theta}}{2}\left[ \left(\epsilon^2\right)'+2\,A'_\rVert\,\epsilon+2\,A_\rVert\,\epsilon'\right]\\
\label{eq:pert113}
\Delta g_{\rVert22}^{(AN)}=&-\sin^2 \theta\,\left(\epsilon^2+2\,A_\rVert\,\epsilon\right).
\end{align}
\end{subequations}
The script "(LTB)" (up or down is the same) means that the quantity refers to Lemaitre-Tolman-Bondi Universe, while "(AN)" refers to anisotropic Universe. 
In other words the metric (\ref{eq:LTBplusAnisotropic2}) is be able to describe the inhomogeneity and axial anisotropy of the Universe, on the other hand a very interesting thing is that it has been decomposed in the sum of a LTB metric with null curve and a a term that contains whole information about the anisotropy, 
$\epsilon (r, t)$.
\\
For completeness reasons, it is possible  to rewrite in the symmetric way the metric as 
\begin{equation}
\label{eq:LTBplusAnisotropic3}
g_{\mu \nu}\equiv g_{\perp \mu \nu}^\text{(LTB)}+\Delta g_{\perp \mu \nu}^{(AN)}
\end{equation}
where $g^{(LTB)}_{\perp \mu \nu}$ is obtained by eq. (\ref{eq:pert10}) with the substitution of $A_{\perp}$ instead of $A_{\rVert}$ and 
\begin{subequations}
\allowdisplaybreaks
\begin{align}
\label{eq:pert211}
\Delta {g_{\perp 11}^{(AN)}}=& \sin^2 \theta \left( \epsilon'^2-2\,A'_\perp\,\epsilon' \right)\\
\label{eq:pert212}
\Delta g_{\perp 12}^{(AN)}=& \frac{\sin{2 \theta}}{2}\left[ \left(\epsilon^2\right)'-2\,A'_\perp\,\epsilon-2\,A_\perp\,\epsilon'\right]\\
\label{eq:pert213}
\Delta g_{\perp 22}^{(AN)}=& \cos^2 \theta \left( \epsilon^2-2\,A_\perp\,\epsilon \right)\\
\label{eq:pert214}
\Delta g_{\perp 33}^{(AN)}=& \sin^2 \theta \left( \epsilon^2-2\,A_\perp\,\epsilon \right).
\end{align}
\end{subequations}
\\
\\
\section{III. Einstein's equations in LTB-Bianchi I Universe}

In this Section we want to write the Einstein's equations taking into account the LTB-Bianchi I metric. To this end we suppose a very small anisotropy of the Universe, in order to have  
\begin{subequations}
\label{eq:limit_condition}
\allowdisplaybreaks
\begin{align}
\epsilon(r,t)&\ll A_\rVert(r,t)\\
\epsilon'(r,t)&\ll A_\rVert'(r,t).
\end{align}
\end{subequations}
These positions permit to expand our results to the first order in $\epsilon$, in other terms:
\begin{equation}
\label{expansion}
\Delta g^{(AN)}_{\mu \nu} \rightarrow \delta g^{(AN)}_{\mu \nu}
\end{equation}
with
\begin{equation}
\label{deltag}
\delta g^{(AN)}_{\rVert 11} = -2 A'_{\rVert} \epsilon' {\text{cos}}^2 \theta 
\end{equation}
\begin{equation}
\delta g^{(AN)}_{\rVert 22} = -2 A_{\rVert} \epsilon {\text{sin}}^2 \theta 
\end{equation}
\begin{equation}
\delta g^{(AN)}_{\rVert 12} = 2 A_{\rVert} \epsilon' \text{sin} \theta \text{ cos} \theta. 
\end{equation}
At this point we can calculate the Christoffel connection  to the first order in $\epsilon$ (we repeats that the script $(LTB)$ and $(AN)$ are indifferently written up or down):
\begin{eqnarray}
\label{christoffel}
\Gamma_{\mu \nu}^{\alpha} & = & \frac {1} {2} g^{\alpha \rho} (\partial_{\mu} \, g_{\nu \rho} + \partial_{\nu} \, g_{\rho \mu} - \partial_{\rho} \, g_{\mu \nu}) \simeq{\phantom {xxxxxxxxxxxxxx}} \nonumber \\
&\simeq & \frac {1} {2} \left( g^{\alpha \rho}_{\small{(LTB)}} + \delta g^{\alpha \rho}_{(AN)} \right) \left[ \partial_{\mu} (g_{\nu \rho}^{(LTB)}+ \delta g^{(AN)}_{\nu \rho})+ \right. 
\nonumber \\
&+& \left. \partial_{\nu} (g_{\rho \mu}^{(LTB)}+ \delta g^{(AN)}_{\rho \mu})-\partial_{\rho} (g_{\mu \nu}^{(LTB)}+ \delta g^{(AN)}_{\mu \nu}) \right], 
\end{eqnarray}
that, neglecting the second order terms in $\epsilon^2$, becomes
\begin{eqnarray}
\label{newgamma}
\Gamma_{\mu \nu}^{\alpha} & \simeq & \frac {1} {2}  g^{\alpha \rho}_{(LTB)} \left( \partial_{\mu} g_{\nu \rho}^{(LTB)} + \partial_{\nu} g_{\rho  \mu}^{(LTB)}- \partial_{\rho} g_{\mu \nu}^{(LTB)} \right) + {\phantom {xxx}} \nonumber \\
&+& \frac {1} {2}  g^{\alpha \rho}_{(LTB)} \left( \partial_{\mu} \delta g_{\nu \rho}^{(AN)} + \partial_{\nu} \delta g_{\rho  \mu}^{(AN)}- \partial_{\rho} \delta g_{\mu \nu}^{(AN)} \right) + \nonumber \\
&+& \frac {1} {2} \delta  g^{\alpha \rho}_{(AN)} \left( \partial_{\mu} g_{\nu \rho}^{(LTB)} + \partial_{\nu} g_{\rho  \mu}^{(LTB)}- \partial_{\rho}  g_{\mu \nu}^{(LTB)} \right).
\end{eqnarray}
The first term in eq.~(\ref{newgamma}) is just $\Gamma_{\mu \nu}^{\alpha (LTB)}$,  the Christoffel connection with the metric tensor $g_{\mu \nu}^{(LTB)}$, and putting
\begin{equation}
\label{sigma}
\Sigma_{\mu \nu}^{\alpha} \equiv \frac {1} {2} {g^{\alpha \rho}}^{(LTB)}( \partial_{\mu} \delta g_{\nu \rho}^{(AN)} + \partial_{\nu} \delta g_{\rho \mu}^{(AN)} - \partial_{\rho} \delta g_{\mu \nu}^{(AN)})
\end{equation}
and
\begin{equation}
\label{theta}
\Theta_{\mu \nu}^{\alpha} \equiv  \frac {1} {2} \delta {g^{\alpha \rho}}^{(AN)} ( \partial_{\mu} g_{\nu \rho}^{(LTB)} + \partial_{\nu} g_{\rho \mu}^{(LTB)} - \partial_{\rho} g_{\mu \nu}^{(LTB)}),
\end{equation}
it is possibile to write the Christoffel connection at the first order as
\begin{equation}
\label{christoffelconnection}
\Gamma_{\mu\nu}^{\alpha} = {\Gamma_{\mu \nu}^{\alpha \, (LTB)}} +\Sigma_{\mu \nu}^{\alpha} + \Theta_{\mu \nu}^{\alpha}\, .
\end{equation}
Let us calculate the Ricci tensor 
\begin{eqnarray}
\label{riccitensor}
R_{\nu \alpha}  & = & \partial_{\mu} \Gamma_{\nu \alpha}^{\mu} - \partial_{\nu} \Gamma_{\mu \alpha}^{\mu} +  \Gamma_{\mu \rho}^{\mu}  \Gamma_{\nu \alpha}^{\rho} - \Gamma_{\nu \rho}^{\mu}  \Gamma_{\mu \alpha}^{\rho}
\end{eqnarray}
with $\Gamma_{\nu \alpha}^{\mu}$ given by eq.(\ref{christoffelconnection}). Therefore we have 
\begin{eqnarray}
\label{ricci_explicate}
R_{\nu \alpha} & = & \partial_{\mu} ( \Gamma_{\nu \alpha}^{\mu \, (LTB)} + \Sigma_{\nu \alpha}^{\mu} + \Theta_{\nu \alpha}^{\mu}) + {\phantom {xxxxxx}}  \nonumber 
\\
& -&
\partial_{\nu} ( \Gamma_{\mu \alpha}^{\mu \, (LTB)} + \Sigma_{\mu \alpha}^{\mu} + \Theta_{\mu \alpha}^{\mu})
+ \nonumber \\
& +& (\Gamma_{\mu \rho}^{\mu \, (LTB)} + \Sigma_{\mu \rho}^{\mu} + \Theta_{\mu \rho}^{\mu} ) 
(\Gamma_{\nu \alpha}^{\rho \, (LTB)} + \Sigma_{\nu \alpha}^{\rho} + \Theta_{\nu \alpha}^{\rho} ) + \nonumber \\
&-& (\Gamma_{\nu \rho}^{\mu \, (LTB)} + \Sigma_{\nu \rho}^{\mu} + \Theta_{\nu \rho}^{\mu} ) 
(\Gamma_{\mu \alpha}^{\rho \, (LTB)} + \Sigma_{\mu \alpha}^{\rho} + \Theta_{\mu \alpha}^{\rho} ). \nonumber \\
\end{eqnarray}
When we multiply in eq.~(\ref{ricci_explicate}),  neglecting the second order terms $\Sigma \Sigma$, $\Theta \Theta$, $\Sigma \Theta$ and $\Theta \Sigma$, and putting 
\begin{eqnarray}
\label{Ricci_LTB}
R_{\nu \alpha}^{(LTB)} & = & \partial_{\mu} \Gamma_{\nu \alpha}^{\mu \, (LTB)} - \partial_{\nu} \Gamma_{\mu \alpha}^{\mu \, (LTB)} + {\phantom {xxxx}} \nonumber \\  
& + & \Gamma_{\mu \rho}^{\mu \, (LTB)}  \Gamma_{\nu \alpha}^{\rho \, (LTB)} - \Gamma_{\nu \rho}^{\mu \, (LTB)}  \Gamma_{\mu \alpha}^{\rho \, (LTB)}  , {\phantom {xxxx}} 
\end{eqnarray}
\begin{eqnarray}
\label{rsigma}
R_{\nu \alpha}^{(\Sigma)} & \equiv & \partial_{\mu} \Sigma_{\nu \alpha}^{\mu} - \partial_{\nu} \Sigma_{\mu \alpha}^{\mu} + \Sigma_{\mu \rho}^{\mu} {\Gamma_{\nu \alpha}^{\rho \, (LTB)}}+ {\Gamma_{\mu \rho}^{\mu \, (LTB)}} \, \Sigma_{\nu \alpha}^{\rho}+\nonumber
\\
& &  - {\Gamma_{\nu \rho}^{\mu \, (LTB)}} \, \Sigma_{\mu \alpha}^{\rho} - \Sigma_{\nu \rho}^{\mu} \, {\Gamma_{\mu \alpha}^{\rho \, (LTB)}}
\end{eqnarray}
\begin{eqnarray}
\label{rtheta}
R_{\nu \alpha}^{(\Theta)} & \equiv & \partial_{\mu} \Theta_{\nu \alpha}^{\mu} - \partial_{\nu} \Theta_{\mu \alpha}^{\mu} + \Theta_{\mu \rho}^{\mu} {\Gamma_{\nu \alpha}^{\rho \, (LTB)}}+ {\Gamma_{\mu \rho}^{\mu \, (LTB)}} \, \Theta_{\nu \alpha}^{\rho}+\nonumber
\\
& &  - {\Gamma_{\nu \rho}^{\mu \, (LTB)}} \, \Theta_{\mu \alpha}^{\rho} - \Theta_{\nu \rho}^{\mu} \, {\Gamma_{\mu \alpha}^{\rho \, (LTB)}},
\end{eqnarray}%
the Ricci tensor to the first order in $\delta$ can be written as
\begin{equation}
\label{ricci}
R_{\nu \alpha} = R_{\nu \alpha}^{(LTB)} + R_{\nu \alpha}^{(\Sigma)} + R_{\nu \alpha}^{(\Theta)}.
\end{equation}
\\
In order to consider the perturbations of the energy momentum tensor, we consider a general anisotropic density energy given by:

\begin{eqnarray}
\label{rhomatt}
\rho_{mat}(r,t, \theta)  &\equiv & \rho_{\rVert mat}(r,t) \, \text{sin}^2 \theta + \rho_{\perp mat}(r,t) \, \text{cos}^2 \theta = \nonumber
\\
&=& \rho_{\rVert mat} (r,t)+ \delta_{mat} (r,t) \text{cos}^2 \theta
\end{eqnarray}
where $\delta_{mat} \equiv \rho_{\perp mat} - \rho_{\rVert mat}$.
The density that we choose has a planar symmetry, because of the consistency with the metric that we are working with. This choice allows us to rewrite the energy momentum tensor and its trace as:
\begin{equation}
\label{TT}
T_{\mu}^{\nu} \equiv T_{\mu}^{\nu \,\,(LTB)} + \Delta T_{\mu}^{\nu}
\end{equation}
\begin{equation}
\label{TTT}
T \equiv T^{(LTB)} + \Delta T
\end{equation}
where, in general
\begin{equation}
T_{\mu}^{\, \nu \, (LTB)}= \text{diag} [\rho(r,t), -p(r,t),-p(r,t),-p(r,t)].
\end{equation}%
However, all these definitions must be view as a first order correction to the usual energy momentum tensor in the LTB case. This point of view becomes clear if we look at the usual perturbation theory provided by \cite{Mukhanov:1990me}. In fact, our particular definition is fully consistent with the Mukhanov's one whether we fix $\delta p = V = \sigma = 0$ in Eq. (5.1) of \cite{Mukhanov:1990me}: this means that we are only considering the perturbation in the energy density and we neglect the effect of a different pressure along two directions (in particular, we continue in using a pressureless matter fluid everywhere). For sure, what we did is a strong constraint. By the way, the particular choice of the perturbation is not relevant for the purposes of this paper.
\\
In order to be explicit we write:
\begin{equation}
\label{energ_impul}
T_{\mu}^{\, \nu} = T_{\mu}^{\, \nu \, (LTB)} + \delta T_{\mu}^{\, \nu \, (AN)}
\end{equation}
and
\begin{eqnarray}
\label{energ_impul2}
T_{\nu \alpha} & = & ( g_{\nu \mu}^{(LTB)} + \delta g_{\nu \mu}^{(AN)}) (T_{\alpha}^{\mu \, (LTB)} + \delta T_{\alpha}^{\mu \, (AN)}) = \nonumber \\
&\simeq& T_{\nu \alpha}^{(LTB)} + \delta g_{\nu \mu}^{(AN)} T_{\alpha}^{\mu (LTB)} + g_{\nu \mu}^{(LTB)} \delta T_{\alpha}^{\mu \, (AN)}. \phantom{xxxxx}
\end{eqnarray}
%
%
As regards the energy conditions, we consider the general energy-momentum tensor:
\begin{equation}
T_{\mu\nu}=\rho\,u_\mu u_\nu+p\left( -g_{\mu\nu}+u_\mu u_\nu \right)
\end{equation}
with $u_\mu u^\mu=1$. Hence, energy conditions state:
\begin{itemize}
\item{weak energy condition: $T_{\mu\nu}u^\mu u^\nu\ge 0$}
\item{dominant energy condition: by defining $W^\mu=T^{\mu\nu}u_\nu$, $W^\mu W_\mu\ge 0$}
\item{strong energy condition: $T_{\mu\nu}u^\mu u^\nu\ge\frac{1}{2}T u^\mu u_\mu$}
\item{null energy condition: $T_{\mu\nu}k^\mu k^\nu\ge 0$, where $k^\mu$ is a light-like vector.}
\end{itemize}

A well-known, the hierarchy among these conditions is the following: strong implies null, dominant implies weak and weak implies null. In this way, by providing that the dominant condition holds, also weak and null are satisfied as well. In particular, for energy-momentum (36), with $p=0$, they become:
\begin{itemize}
\item{weak: $\rho_{mat}\ge 0$}
\item{dominant: $\rho_{mat}^2\ge 0$}
\item{strong: $\rho_{mat}\ge0$}
\item{null: $\rho_{mat}\ge 0$.}
\end{itemize}

%
%
%

%
The Einstein's equations in this Universe are:
\begin{eqnarray}
\label{einstein}
& & 
R_{\mu}^{\, \nu (LTB)}+R_{\mu \nu}^{(LTB)} \delta g^{\alpha \nu}+ (R_{\mu \alpha}^{(\Sigma)} + R_{\mu \alpha}^{\Theta}) {g^{\nu \alpha}}^{(LTB)} = \nonumber
\\
& & 
8 \pi G [ T_{\mu}^{\, \nu \, (LTB)} + \delta T_{\mu}^{\, \nu \, (AN)} + \frac {1} {2} \delta_{\mu}^{\, \nu} (T^{(LTB)} + \delta T^{(AN)})]. \nonumber
\\
& & 
\end{eqnarray}
{\section {IV. luminosity distance}}
The concept of distance depends on the assumed model of the Universe and on the matter distribution in it. The measured distance are influenced by inhomogeneities and anisotropy of the Universe, see for example \cite{kristian} and  \cite{durrer}.

The luminosity distance is one of the most important quantity to understand the presence of dark energy in the Universe, considering the photon coming from Supernovae Ia.
In this section we want to calculate the luminosity distance for our metric eq.(\ref{eq:LTBplusAnisotropic2}). The reciprocity theorem by Etherington (1993) \cite{etherington} and popolarized by Ellis \cite{ellis} connects the angular diameter distance $d_A$ and the luminosity distance $d_L$ by
\begin{equation}
\label{dist}
d_L= (1+z)^2 d_A
\end{equation}
where
\begin{equation}
\label{dist2}
d(\text{ln} \, d_A) = \frac {1} {2} \nabla_{\alpha} p^{\alpha}  d \tau
\end{equation}
with $\tau$ temporal affine parameter and $p^{\alpha}=dx^{\alpha}/d \tau$ quadri-momentum of a generic signal that is started from the Supernova and reaches us. To our end it is necessary to calculate $\nabla_\alpha p^\alpha\equiv \partial_\alpha p^\alpha+{\Gamma_{\alpha \mu}}^\alpha p^\mu$:
\begin{multline}
\label{eq:anisotropicdA1}
\nabla_\alpha p^\alpha =\partial_\alpha p^\alpha+{\Gamma_{\alpha \mu}}^\alpha p^\mu=\partial_\alpha p^\alpha+\frac{\partial_\mu \sqrt{-g}}{\sqrt{-g}}\,p^\mu=\\=\partial_0 p^0+\partial_1 p^1+\frac{\partial_0 \sqrt{-g}}{\sqrt{-g}}\,p^0+\frac{\partial_1 \sqrt{-g}}{\sqrt{-g}}\,p^1
\end{multline}
with
\begin{equation}
\label{detg}
-g =  A^2_{\rVert} \,  (g_{11} \, g_{22} - g_{12}^2) \, {\text{sin}}^2 \theta \equiv A^2_{\rVert} \,B^2(t,r,\theta)\,  {\text{sin}}^2 \theta \, 
\end{equation}
where we define
\begin{equation}
\label{def_B}
B(t,r,\theta) \equiv \sqrt{g_{11}(t,r,\theta) \, g_{22}(t,r,\theta) - g_{12}^2(t,r,\theta)  }.
\end{equation}
In this way we obtain:
\begin{subequations}
\allowdisplaybreaks
\begin{align}
\frac{\partial_0\sqrt{-g}}{\sqrt{-g}}=\frac{\left(\partial_0A_\rVert B+A_\rVert \partial_0 B\right) \sin \theta}{A_\rVert B\sin \theta}=\frac{\partial_0 A_\rVert}{A_\rVert}+\frac{\partial_0 B}{B}\\
\frac{\partial_1\sqrt{-g}}{\sqrt{-g}}=\frac{\left(\partial_1A_\rVert B+A_\rVert \partial_1 B\right) \sin \theta}{A_\rVert B\sin \theta}=\frac{\partial_1 A_\rVert}{A_\rVert}+\frac{\partial_1 B}{B}.
\end{align}
\end{subequations}
This permits to write eq.(\ref{eq:anisotropicdA1}) as
\begin{multline}
\label{eq:anisotropicdA2}
\partial_0 p^0+\partial_1 p^1+\left( \frac{\partial_0A_\rVert}{A_\rVert}+\frac{\partial_0B}{B} \right)p^0+\left( \frac{\partial_1A_\rVert}{A_\rVert}+\frac{\partial_1B}{B} \right)p^1=\\=\partial_0 p^0+\partial_1 p^1+\frac{1}{A_\rVert}\frac{d A_\rVert}{d\tau}+\left( \frac{\partial_0 B}{B}\,p^0+\frac{\partial_1 B}{B}\,p^1 \right).
\end{multline}
In the last equation we have considered the general relation between partial derivates in the coordinates $x^{\alpha}$ and total derivates in the affine time $\tau$. In fact if $\Phi$ is a generic function that depends from the coordinates it is possible to write:
\begin{equation}
\label{eq:usefullRelation}
\frac{d\,\Phi(x^\alpha(\tau))}{d\tau}=\frac{\partial\,\Phi(x^\alpha)}{\partial x^\beta}\,\frac{d x^\beta}{d\tau}\equiv \partial_\beta \Phi\,p^\beta.
\end{equation}  
On the other hand as regards $B$, we must write $
\frac{dB}{d\tau}=\partial_0B\,p^0+\partial_1B\,p^1+\partial_2B\,p^2,
$
but we are considering radial signal, therefore $p^2 \equiv d \theta/d \tau$, 
that is to say $ \theta (\tau)=cost.$ In this way we can consider $\theta$ as a parameter that is  able to locate the trajectory of propagation of light. This employment permits to write:
\begin{equation}
B(r,t,\theta)\approx B(r(\tau),t(\tau),\theta) \Rightarrow \frac{d B}{d\tau}\approx \partial_0 B p^0+\partial_1 B p^1.
\end{equation}
Therefore, eq.~\eqref{eq:anisotropicdA1} becomes:
\begin{equation}
\label{nabla}
\nabla_\alpha p^\alpha=\partial_0p^0+\partial_1p^1+\frac{1}{A_\rVert}\frac{d A_\rVert}{d\tau}+\frac{1}{B}\frac{d B}{d\tau}.
\end{equation}
As regards the partial derivative of $p^{\alpha}$, remembering that we are considering the radial propagation of signals, the relevant components are:
\begin{subequations}
\allowdisplaybreaks
\begin{align}
dp^0+{\Gamma_{00}}^0 dx^0 p^0+{\Gamma_{10}}^0\left( dx^1 p^0+dx^0 p^1\right)+{\Gamma_{11}}^0 dx^1 p^1=0\\
dp^1+{\Gamma_{00}}^1 dx^0 p^0+{\Gamma_{10}}^1\left( dx^1 p^0+dx^0 p^1\right)+{\Gamma_{11}}^1 dx^1 p^1=0
\end{align}
\end{subequations}
from which we have:
\begin{subequations}
\allowdisplaybreaks
\begin{align}
\partial_0p^0=-\left( {\Gamma_{00}}^0 p^0+{\Gamma_{10}}^0 p^1 \right) 
\phantom{.} \\
 \partial_1p^1=-\left( {\Gamma_{10}}^1 p^0+{\Gamma_{11}}^1 p^1 \right).
\end{align}
\end{subequations}
In order to complete the analysis observe that ${\Gamma_{00}}^0={\Gamma_{10}}^0=0$ and
\begin{multline}
{\Gamma_{10}}^1 =\frac{1}{2}g^{11}\left( \partial_1g_{01}+\partial_0g_{11}-\partial_1g_{10} \right) +   \\
+ 
\frac{1}{2}g^{12}\left( \partial_1 g_{02}+\partial_0g_{21}-\partial_2g_{10} \right)=  \\
= -\left( g^{11}X\partial_0X+g^{12}F \partial_0F \right) \phantom{xxxxxxxxxx}
\end{multline}
\begin{multline}
{\Gamma_{11}}^1=\frac{1}{2}g^{11}\left( \partial_1g_{11}+\partial_1g_{11}-\partial_1g_{11} \right)+ \\+\frac{1}{2}g^{12}\left( \partial_1g_{12}+\partial_1g_{21}-\partial_2g_{11} \right)=
\\=
-\left( g^{11}X\partial_1X+2g^{12}F\partial_1F-X\partial_2X \right) \phantom{xx}
\end{multline}
where we have put
\begin{equation}
\label{position}
g_{11} \equiv X^2 \;\;\;\; g_{22} \equiv Y^2 \;\;\;\; g_{12} \equiv F^2.
\end{equation}
\\
Now we work in small approximation of anisotropy, in order to use eq.(\ref{christoffelconnection}) to the lower order, in this way it is possibile to write:
\begin{subequations}
\allowdisplaybreaks
\begin{align}
&{\Gamma_{10}}^1\rightarrow{\Gamma_{10}^{\phantom{x} 1 \text{(LTB)}}}=\frac{\partial_0 \partial_1A_\rVert}{\partial_1A_\rVert}\\
&{\Gamma_{11}}^1\rightarrow{\Gamma_{11}^{\phantom{x} 1\text{(LTB)}}}=\frac{\partial_1^2A_\rVert}{\partial_1A_\rVert}
\end{align}
\end{subequations}
where $\partial_1^2\equiv \frac{\partial^2}{\partial r^2}$. Therefore eq.(\ref{nabla}) is
\begin{equation}
\label{eq:appr1}
\nabla_\alpha p^\alpha\approx-\left( \frac{\partial_0 \partial_1A_\rVert}{\partial_1A_\rVert}\,p^0+\frac{\partial_1^2A_\rVert}{\partial_1A_\rVert}\,p^1 \right)+\frac{1}{A_\rVert} \frac{dA_\rVert}{d\tau}+\frac{1}{B} \frac{dB}{d\tau}.
\end{equation}
Taking into account eq.(\ref{eq:usefullRelation}) it is possible to write the first two terms in eq.(\ref{eq:appr1}) as $\frac{1}{\partial_1 A_\rVert}\frac{dA_\rVert}{d\tau}$. 
\\
Inserting eq.(\ref{eq:appr1}) in eq. (\ref{dist2}) it is possible to obtain $d_A$, in fact we have:
\begin{equation}
\frac{dd_A}{d_A}\approx \frac{1}{2}\left( \frac{1}{A_\rVert}\frac{dA_\rVert}{d\tau}+\frac{1}{B}\frac{dB}{d\tau}-\frac{1}{\partial_1A_\rVert}\frac{d \partial_1A_\rVert}{d\tau} \right)d\tau
\end{equation}
and integrating in $\tau$ we obtain:
\begin{equation}
\label{eq:aniLTBdA}
d_A(r,t,\theta)=\sqrt{\frac{A_\rVert(r,t)B(r,t,\theta)}{\partial_1A_\rVert(r,t)}}.
\end{equation}
This expression of the luminosity distance reduces to the isotropic limit of the LTB metric, in fact we have
\begin{equation}
X(r,t,\theta)\rightarrow \partial_1 A_\rVert(r,t) \,\,\,\, A(r,t,\theta)\rightarrow A_\rVert(r,t) \,\,\,\, F(r,t,\theta)\rightarrow0,
\end{equation}
therefore we obtain the limit
\begin{multline}
B(r,t,\theta) \rightarrow A_\rVert(r,t)\,\partial_1 A_\rVert(r,t)\Rightarrow\\ \Rightarrow d_A(r,t,\theta)\rightarrow d_A^\text{(LTB)}(r,t)=A_\rVert(r,t).
\end{multline}
\\
\\
\section{V. Relation between coordinates and redshift}
In this section we want to calculate the luminosity distance in order to obtain an operative expression and therefore to apply it to experimental data. The eq. (\ref{eq:aniLTBdA}) give us the luminosity distance:
\begin{equation}
\label{eq:aniLTBdL}
d_L=(1+z)^2 \sqrt{\frac{A_\rVert(r,t) B(r,t,\theta)}{\partial_1A_\rVert(r,t)}},
\end{equation}
but this expression is not directly applicable, because of it depends on $(r,t)$ coordinates and on the redshift. Theferore it is necessary to find the relations
$r(z)$ e $t(z)$. To this end let us consider the definition of redshift and let us use the static observators classes that are geodetic also ( ${\Gamma_{00}}^\mu=0$). We consider light signal, that is to say $p^0\propto 1/\delta t$, in this way we write:
\begin{equation}
1+z\equiv\frac{(g_{\mu \nu}u^\mu p^\nu)_\text{em}}{(g_{\mu \nu}u^\mu p^\nu)_\text{oss}}=\frac{p^0_\text{em}}{p^0_\text{oss}}= \frac{\delta t_\text{oss}}{\delta t_\text{em}}\Rightarrow1+z(\tau)=\frac{\delta t_\text{oss}}{\delta t(\tau)}
\end{equation}
where $u^\mu=(1,0,0,0)$ e $g_{00}=1$. Now we derive with respect to $\tau$ and we have
\begin{multline}
\label{eq:dtdtau1}
\frac{dz(\tau)}{d\tau}=-\frac{\delta t_\text{oss}}{\delta t(\tau)^2}\frac{d \delta t(\tau)}{d\tau}\equiv -\frac{1+z(\tau)}{\delta t(\tau)}\frac{d \delta t(\tau)}{d \tau}\Rightarrow \\ \Rightarrow \frac{d \delta t}{d \tau}=-\frac{\delta t}{1+z} \frac{dz}{d\tau}.
\end{multline}
On the other hand for geodetic radial signals we have $ds^2=0$ e $d\theta=d\phi=0$ that gives
\begin{equation}
\label{eq:aniLightPropagation}
dt^2-X(r,t,\theta)^2dr^2=0\Rightarrow dt=\pm X(r,t,\theta)dr.
\end{equation}
As regards the ambiguity of the sign  we must consider the minus sign because of increasing the distance ($dr >0$) we have a more ancient signal ($dt<0$). When we consider the signals that respectively start at time $t$ and $t+\delta t$, the eq. (\ref{eq:aniLightPropagation}) must be valid, therefore we have:
\begin{subequations}
\allowdisplaybreaks
\begin{align}
\label{eq:aniGeodesic1}
\frac{dt}{d\tau}&=-X(r,t,\theta) \frac{dr}{d\tau}\\
\label{eq:aniGeodesic2}
\frac{d\left(t+\delta t\right)}{d\tau}&=-X(r,t+\delta t,\theta) \frac{dr}{d\tau}.
\end{align}
\end{subequations}
Eq.(\ref{eq:aniGeodesic2}) can be written as
\begin{equation}
\frac{dt}{d\tau}+\frac{d \delta t}{d \tau}\approx -\left[ X(r,t,\theta)+\delta t\,\partial_0 X(r,t,\theta) \right]\frac{dr}{d\tau}
\end{equation}
that, taking into account eq.(\ref{eq:aniGeodesic1}), can be written as
\begin{equation}
\label{eq:dtdtau2}
\frac{d \delta t}{d \tau}\approx -\delta t\,\partial_0X(r,t,\theta)\frac{dr}{d\tau}= -\delta t\,\partial_0X(r,t,\theta)\frac{dr}{dz}\frac{dz}{d\tau}.
\end{equation}
Now eqs. (\ref{eq:dtdtau1}) and (\ref{eq:dtdtau2}) are equal, therefore we have 
\begin{equation}
\frac {dr} {dz} = \frac{1}{1+z} \,\frac{1}{\partial_0 X(r,t,\theta)}.
\end{equation}
As regards $t(z)$ it is important to remember that 
\begin{equation}
\frac{dt}{d\tau}=\frac{dt}{dz}\frac{dz}{d\tau}\qquad\text{e}\qquad\frac{dr}{d\tau}=\frac{dr}{dz}\frac{dz}{d\tau}
\end{equation}
in this way taking into account eq. (\ref{eq:aniGeodesic2}) we obtain the relation:
\begin{equation}
\frac{dt}{dz}=-\frac{1}{1+z}\,\frac{X(r,t,\theta)}{\partial_0 X(r,t,\theta)}.
\end{equation}
Putting all togheter, we are be able to write the luminosity distance as a function of the redshift $z$ and the angle $\theta$ 
\begin{subequations}
\begin{align}
\label{eq:dLvszandtheta}
d_L(z,\theta)=(1+z)^2&\left[ \frac{A_\rVert(r_\theta(z),t_\theta(z))}{\partial_1A_\rVert(r_\theta(z),t_\theta(z))}\,B(r_\theta(z),t_\theta(z),\theta)^\frac{1}{2} \right]^\frac{1}{2} \phantom{xxxxxxxxx}\\
\label{eq:rthetavsz}
\frac{dr_\theta(z)}{dz}=&\frac{1}{1+z}\; \frac{1}{\partial_0 X(r_\theta(z),t_\theta(z),\theta)}\\
\label{eq:tthetavsz}
\frac{dt_\theta(z)}{dz}=&-\frac{1}{1+z} \; \frac{X(r_\theta(z),t_\theta(z),\theta)}{\partial_0X(r_\theta(z),t_\theta(z),\theta)}.
\end{align}
\end{subequations}
It is important to observe that subscript $\theta$  remember us that the functions $r$ and $t$ are determined by the resolution of the system given by eqs (\ref{eq:rthetavsz}) and (\ref{eq:tthetavsz}), where the angle $\theta$ is fixed and considered as a constant parameter during the propagation of light. 
\\
\\
\section{VI. Comparison with experimental data}
The accelerating expansion of the universe is driven by mysterious energy with negative pressure known as Dark Energy. In spite of all the observational evidences, the nature of Dark Energy is still a challenging problem in theoretical physics,
therefore there has been a new interest in studying alternative cosmological models \cite{krasinsky}.  
\\
In the context of FLRW models the acceleration of the Universe requires the presence of a cosmological constant. But it does not appear to be natural to introduce the presence of a cosmological constant and does not appear to be natural to introduce the dark energy. 
\\
In this Section we consider the comparison between experimental data, in particular with Union 2 data set of Supernovae Ia and our inhomogeneous and anisotropic  Universe. Let us suppose a small anisotropy in order to write the functions $A_\rVert$ e $A_\perp$ as  solutions of a LTB Universe with null curvature and matter dominated. We have
\begin{subequations}
\begin{align}
A_\rVert(r,t)=&r\left( 1+\frac{3}{2}H_{\rVert}(r)\,t \right)^\frac{2}{3}\\
A_\perp(r,t)=&r\left( 1+\frac{3}{2}H_{\perp}(r)\,t \right)^\frac{2}{3}
\end{align}
\end{subequations}
where we have considered the following parametrization
\begin{equation}
H_{\rVert/\perp}(r)=H_{\rVert/\perp}+\Delta H_{\rVert/\perp}\exp\left({-\frac{r}{r_{\rVert/\perp}}}\right).
\end{equation}
In this way we have the possibility to obtain again the simple model in which  
$H_\rVert=H_\perp$, $\Delta H_\rVert=\Delta H_\perp$ and $r_\rVert=r_\perp$. 
Let us consider that today and in our position in the Universe ($t=0$ e $r=0$) the Hubble constant is $67.3 \pm 1.2\,\frac{\text{km}}{s}/\text{Mpc}$ 
\cite{Planck}.  We have the following conditions  $H_\rVert+\Delta H_\rVert=H_\perp+\Delta H_\perp=67.3$, therefore we have:
\begin{subequations}
\begin{align}
H_\rVert=&\,67.3-\Delta H_\rVert\\
H_\perp=&\,67.3-\Delta H_\perp.
\end{align}
\end{subequations}
In this way we have not the six parameters of the model, now they are four:  
$\Delta H_\rVert,\,\Delta H_\perp,\,r_\rVert$ e $r_\perp$.
At this point we  remember the limits given by \eqref{eq:limit_condition}, therefore for $\epsilon \sim 0$ we have
\begin{equation}
\label{eq:eps_null_limit}
A_\perp (r,t) \approx A_\rVert (r,t)\Rightarrow H_\rVert(r) \approx H_\perp (r).
\end{equation}
This condition must be transfered to the four parameters.
\\
\\
\\
The condition eq. \eqref{eq:eps_null_limit} is obtained when  $\alpha \sim 1$ and $\omega \sim 1$. On the other hand it also must be  $\epsilon'\sim 0$. Therefore we have:
\begin{multline}
A'_{\rVert/\perp}=\left( 1+\frac{2}{2}H_{\rVert/\perp}\,t \right)^\frac{2}{3}+\frac{r\,H'_{\rVert/\perp}\,t}{\left( 1+\frac{3}{2}H_{\rVert/\perp}\,t \right)^\frac{1}{3}}= \\
=
\frac{A_{\rVert/\perp}}{r}+\frac{r^\frac{3}{2}\,H'_{\rVert/\perp}\,t}{A_{\rVert/\perp}}, \phantom{xxxxxxxxxxxxxxx}
\end{multline}
from which we obtain
\begin{subequations}
\begin{align}
\alpha\equiv&\,\frac{r_\perp}{r_\rVert}\\
\omega\equiv&\,\frac{\Delta H_\perp}{\Delta H_\rVert}.
\end{align}
\end{subequations}
\begin{multline}
A'_\rVert-A'_\perp=\frac{A_\rVert-A_\perp}{r}+r^\frac{3}{2}\,t\left( \frac{H'_\perp}{A_\perp}-\frac{H'_\rVert}{A_\rVert} \right)=\\=\frac{\epsilon}{r}+r^\frac{3}{2}\,t\left( -\frac{\Delta H_\perp}{A_\perp r_\perp}+\frac{\Delta H_\rVert}{A_\rVert r_\rVert} \right). \phantom{xxxxx}
\end{multline}
In this way, for $\epsilon \sim 0$ and $\epsilon'\sim 0$, we must write:
\begin{equation}
\frac{\Delta H_\perp}{\Delta H_\rVert}=\frac{A_\perp}{A_\rVert}\frac{r_\perp}{r_\rVert}\Rightarrow \omega=\frac{A_\perp}{A_\rVert}\,\alpha.
\end{equation}
\begin{figure}[h!]
\centering
\includegraphics[scale=0.64]{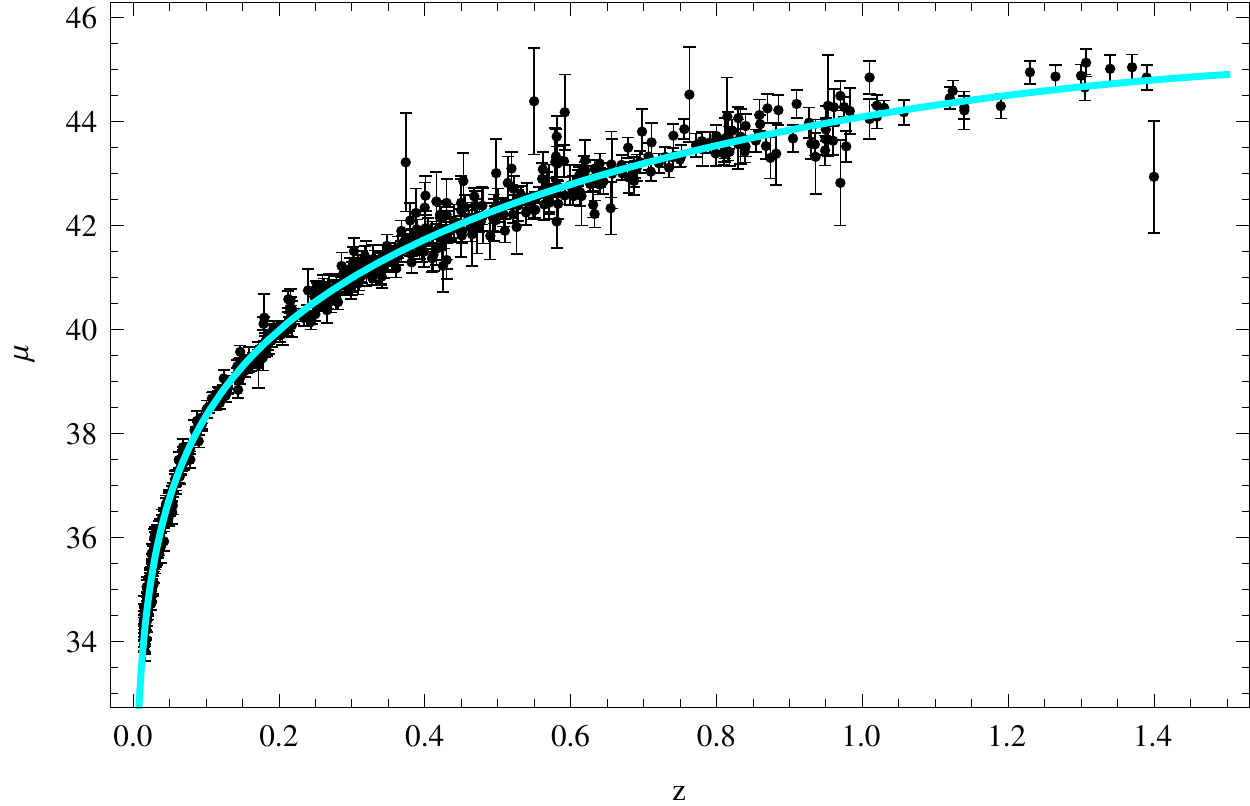}
\caption{Hubble diagram for type Ia supernovae by UNION 2 catalog. The curve is the best fit.}
\label{fig:}
\end{figure}
Therefore $A_\perp/A_\rVert \approx 1$, from which we obtain $\omega \approx \alpha$. In conclusion  we have also three parameters: $\Delta H_\rVert,\, r_\rVert$ and $\alpha$. 
The advantage of this parametrization is that we can change $\Delta H_\rVert$ and $r_\rVert$ as we want, taking into account that  $\alpha \simeq 1$. In our work we have changed $\alpha$ in the range $[0.9,1]$. 
\begin{center}
\begin{tabular}{|c|c|c|}
\hline
$\qquad\Delta H_\rVert\,\left[\text{km}\text{(s Mpc)}^{-1}\right]\qquad$ & $\qquad r_\rVert\,\left[ \text{Gpc} \right]\qquad$ & $\qquad\alpha\qquad$\\
\hline
\hline
25.4&2.88&1.1\\
\hline
\end{tabular}
\end{center}
In table we have the best fit values of the parameters for ${\tilde{\chi}}^2=0.95$. In fig. 1 we have the Hubble diagram for the 557 Supernovae Ia of the UNION 2 catalog. The best fit curve is in the same diagram.  
The fit of the cosmological observational data is in very good agreement, without using any dark energy!

According to our ansatz, it is important to stress that $\rho_{mat} =\rho_{\rVert mat}+\delta_{mat} \cos^2\theta$, so dominant energy condition up to first order gives:
\begin{equation}
\label{dominant}
\delta_{mat}\cos^2\theta\ge-\frac{\rho_{\rVert mat}}{2}
\end{equation}
while the other ones give:
\begin{equation}
\delta_{mat}\cos^2\theta\ge-\rho_{\rVert mat}.
\end{equation}
\\
Furthermore, $\delta_{mat}=\rho_{\perp mat}-\rho_{\rVert mat}$ so, from the Einstein's equation, we have that:
\begin{align}
\rho_{\rVert mat}&=\frac{1}{8\pi\,G}\left[ \left(\frac{\dot A_\rVert}{A_\rVert}\right)^2+2\frac{\dot A_\rVert}{A_\rVert} \frac{\dot A'_\rVert}{A'_\rVert} \right]\nonumber\\
\delta_{mat}&=\frac{1}{8\pi\,G}\left[ \left(\frac{\dot A_\perp}{A_\perp}\right)^2+2\frac{\dot A_\perp}{A_\perp} \frac{\dot A'_\perp}{A'_\perp}-\left(\frac{\dot A_\rVert}{A_\rVert}\right)^2-2\frac{\dot A_\rVert}{A_\rVert} \frac{\dot A'_\rVert}{A'_\rVert} \right].
\end{align}
\\
Hence we have from solutions (70a) and (70b), with conditions (75a), (75b) and (77), that eq.~\eqref{dominant} gives a constraint on parameters which must be satisfied. In particular, by using the best-fit values for $r_\rVert$ and $\Delta H_\rVert$, dominant energy condition requires $\alpha< 1.5$, which is in fully agreement with our analysis.
\\
\\
\section{VII. Conclusion}
In the present paper we have studied the possible effects of an anisotropy and inhomogeneity in the expansion of the Universe. 
\\
The motivation behind this choise is that singly, inhomogeneous cosmological models and Bianchi I cosmological model of the Universe have motivations of thruth that must not be left out for one of the the two models. Both models may be unified in a anisotropic expansion of the inhomogeneous Universe. The LTB-Bianchi I model posseses important specific properties and at the same time this is not too complicated from a physical and mathematical point of view.
\\
In particular we have connected this model on present-day observations as luminosity distance of the Supernovae Ia. We fit observational data from UNION 2 catalog of the Supernovae Ia with a LTB-Bianchi I model of the Universe. The agreement is good. We have not any dark energy in this model.
\\
\\
We are sure that the voids in the Universe dominate, while matter is distributed in a filamentary structure. Therefore photons must travel through the voids and the presence of inhomogeneities can alter the observable with respect to the corresponding  FLRW model of Universe, homogeneous and isotropic.
\\
The key point is that in this model we have two contributions to the Hubble diagram of the Supernovae Ia: inhomogeneity to the large scale geometry and anisotropy can generate dynamically effects that may remove the need for the postulate of dark energy. 
\\
This model must be intended as a first step towards a most general case. The model is oversimplifying for different reasons. First, we have considered only the first order in $\epsilon$. Second, it is necessary to generalize this paper,  a very interesting open question that we will study in future, is to obtain how to treat light-cone average in more realistic cosmological calculation. Third we have considered he simple LTB model of the Universe, but may be very interesting to study more completed inhomogeneous model as for example Swiss-cheese model. Note, finally that several possibility are allowed by our model, it will be interesting to compare this model with other experimental cosmological data.
\\
In future we will study the possibility that inhomogeneity and anisotropy can have significant effects on the propagation of light, with potentially very important effects on cosmological observations and we want to study different observational tests that may confirm this model of the Universe.
\\
\\
\\
\\
The authors would like to thank  M. Gasperini for useful discussions.
This work is supported by the research grant "Theoretical Astroparticle 
Physics" number 2012CPPYP7 under the program PRIN 2012 funded by the Ministero 
dell'Istruzione, Universit\`a e della Ricerca (MIUR). 
%


\begin{thebibliography}{99}
%
%
\bibitem{Ehlers:1966ad} 
  J.~Ehlers, P.~Geren and R.~K.~Sachs,
  J.\ Math.\ Phys.\  {\bf 9}, 1344 (1968).
%
%
\bibitem{Clarkson:1999yj} 
  C.~A.~Clarkson and R.~Barrett,
  Class.\ Quant.\ Grav.\  {\bf 16}, 3781 (1999).
%
%
\bibitem{Clarkson:2010uz} 
  C.~Clarkson and R.~Maartens,
  Class.\ Quant.\ Grav.\  {\bf 27}, 124008 (2010).
%
%
\bibitem{percival} W. Percival, Mon Not. R. Astron. Soc. {\bf 381}, 1053 (2010).
%
%
\bibitem{devega} H.J. de Vega, M.C. Falvella and N.G. Sanchez (2010) ArXiv:1009.3494.
%
%
\bibitem{blanchard} A. Blanchard, Astron. Astrophys. Rev. {\bf18}, 595 (2010).
%
%
\bibitem{clifton} T. Clifton, P.G. Ferreira, A. Padilla and C. Skordis (2011) arXiv:1106.2476.
%
%
\bibitem{copeland} E.L. Copeland, M. Sami and S. Tsujikawa, Int. J. Mod. Phys. D {\bf 15}, 1753 (2006).
%
%
\bibitem{Tauber:1961lbq} 
  G.~E.~Tauber and J.~W.~Weinberg,
  Phys.\ Rev.\  {\bf 122}, no. 4, 1342 (1961).
%
\bibitem{sylos1} 
  F.~S.~Labini,
  Class.\ Quant.\ Grav.\  {\bf 28}, 164003 (2011).
%
%
\bibitem{lemaitre} G. Lemaitre, Annal Soc. Sci Bruxelles {\bf 53}, 51 (1933); Gen Rel. Grav. {\bf 29}, 641 (1997). 
%
%
\bibitem{tolman} R.C. Tolman, Proc. National Acad. Sci. {\bf 20}, 169 (1934).
%
%
\bibitem{bondi} H. Bondi, Mon. Not. Roy. Astron. Soc. {\bf 107}, 410 (1947).
%
%
\bibitem{krasinsky} See for example A. Krasinsky, {\it Inhomogeneous Cosmological Models}, Cambridge University Press (2006).
%
\bibitem{maartens2011} R. Maartens, Phil. Trans. R. Soc. A {\bf 369}, 5155 (2011).
%
%
\bibitem{russel} E. Russel, C.B. Kilinc and O. Pashev, arXiv:1312.3502.
%
%
\bibitem{gurzadyan1} V.G. Gurzadyan {\it et al.}, A\&A {\bf 490}, 929 (2008).
%
\bibitem{gurzadyan2} V.G. Gurzadyan {\it et al}., A\&A {\bf 498}, L1 (2009).
%
\bibitem{bennet} C.L. Bennet, R.S. Hill, G. Hinshaw, M.R. Nolta, N. Odegard, L. Page, D.N. Spergel, J.L. Weiland, Ap.JS {\bf 148}, 97 (2003). 
%
%
%
\bibitem{land}  K.~Land and J.~Magueijo,
  Phys.\ Rev.\ Lett.\  {\bf 95}, 071301 (2005).
%
%
\bibitem{Ralston:2003pf} 
  J.~P.~Ralston and P.~Jain,
  Int.\ J.\ Mod.\ Phys.\ D {\bf 13}, 1857 (2004).
 %
\bibitem{de OliveiraCosta:2003pu} 
  A.~de Oliveira-Costa, M.~Tegmark, M.~Zaldarriaga and A.~Hamilton,
  Phys.\ Rev.\ D {\bf 69}, 063516 (2004).
%
\bibitem{Copi:2003kt} 
  C.~J.~Copi, D.~Huterer and G.~D.~Starkman,
  Phys.\ Rev.\ D {\bf 70}, 043515 (2004).
%
\bibitem{Eriksen:2003db} 
  H.~K.~Eriksen, F.~K.~Hansen, A.~J.~Banday, K.~M.~Gorski and P.~B.~Lilje,
  Astrophys.\ J.\  {\bf 605}, 14 (2004)
  [Erratum-ibid.\  {\bf 609}, 1198 (2004)].
%
%
\bibitem{Hansen:2004vq} 
  F.~K.~Hansen, A.~J.~Banday and K.~M.~Gorski,
  Mon.\ Not.\ Roy.\ Astron.\ Soc.\  {\bf 354}, 641 (2004).
%
\bibitem{Vielva:2003et} 
  P.~Vielva, E.~Martinez-Gonzalez, R.~B.~Barreiro, J.~L.~Sanz and L.~Cayon,
  Astrophys.\ J.\  {\bf 609}, 22 (2004).
%
\bibitem{barrow} J.D. Barrow Phys. Rev. D {\bf 55}, 7451 (1997).  
%
\bibitem{berera} A. Berera, R.V. Buniy and T.W. Kephart, J. Cosmolog. Astropart. Phys. {\bf 0410}, 016 (2004).
%
\bibitem{campanelli1} L. Campanelli, P. Cea and L. Tedesco, Phys. Rev. Lett. {\bf 97}, 131302 (2006); [Erratum-{\it ibid}. {\bf 97} 131302 (2006).
%
\bibitem{campanelli2} L. Campanelli, P. Cea and L. Tedesco, Phys. Rev. D {\bf 76} 063007 (2007).
%
\bibitem{martinez} E. Martinez-Gonzalez, J.L. Sanz, A \& A  {\bf 300}, 346 (1995).
%
%
\bibitem{maartellisstoe1} R. Maartens, G.F.R. Ellis, W. Stoeger, Phys. Rev. D {\bf 51}, 1525 (1995).
%
\bibitem{maartellisstoe2} R. Maartens, G.F.R. Ellis, W. Stoeger, Phys. Rev. D {\bf 51}, 5942 (1995).
%
%
\bibitem{shogin} D.~Shogin and S.~Hervik,
  JCAP {\bf 1310}, 005 (2013).
%
%
\bibitem{zeld} Y.B.  Zel'dovich, Sov. Astron. AJ {\bf 8}, 13 (1964). 
%
%
\bibitem{bertotti} B. Bertotti, Proc. R. Soc. A{\bf 294}, 195 (1966).
%
%
\bibitem{dash} V.M. Dashevskii and V.I. Slysh, Sov. Astron.{\bf 9}, 671 (1966)
%
%
\bibitem{gunn} E.J. Gunn, Astrophys. J. {\bf 150}, 737 (1967). 
%
%
\bibitem{kant} R. Kantowski, Astrophys. J. {\bf 155}, 89 (1969). 
%
%
\bibitem{dyer} C.C. Dyer and R. Roeder, Astrophys. J. {\bf 174}, L115 (1972). 
%
%
\bibitem{weinberg} S. Weinberg, Astrophys. J. {\bf 208} L1 (1976).
%
%
\bibitem{ellis1} G.F. Ellis, B.A. Bassett and P.K.S. Dunsby, Class.Quantum Grav. {\bf 15}, 2345 (1998).
%
%
\bibitem{kristian} J. Kristian and R.K. Sachs, Astrophys. J. {\bf 143}, 379 (1966).
%
%
\bibitem{durrer} 
  J.~Adamek, E.~Di Dio, R.~Durrer and M.~Kunz,
  Phys.\ Rev.\ D {\bf 89}, 063543 (2014).
%
%
\bibitem{etherington} I.M.H. Etheringotn, Philosophical Magazine ser. 7, vol 15, 761 (1933).
%
\bibitem{ellis} G. Ellis, Gen. Rel. Grav. {\bf {39}} 1047 (2007).
%
\bibitem{Mukhanov:1990me}
  V.~F.~Mukhanov, H.~A.~Feldman and R.~H.~Brandenberger,
  Phys.\ Rept.\  {\bf 215} (1992) 203.
%
%
\bibitem{Planck} P.A.R. Ade {\it et al.} [Planck collaboration] Astron. Atsrophys. (2014).
%
\end{thebibliography}
 \end{document}